\documentclass[lnbip]{svmultln}
\usepackage{makeidx}  
\usepackage{graphicx}
\usepackage{amssymb}  
\usepackage{array}
\usepackage{url}

\newcommand{\mypar}[1]{\smallskip\vspace{0.13em}\noindent\textbf{#1.}}

\begin{document}
%
\mainmatter
\title{Integrating Wearable Data into Process Mining: Event, Case and Activity Enrichment\thanks{Accepted manuscript on August 22, 2025, to the 1st International Workshop on Personal and Human-Centric Process Mining (PHPM 2025), held in conjunction with the 7th International Conference on Process Mining (ICPM 2025).}}

\titlerunning{Integrating Wearable Data into Process Mining}
%
\author{Vinicius~Stein~Dani \and Xixi~Lu \and Iris~Beerepoot}
\authorrunning{V. Stein Dani et al.}
%
\institute{Utrecht University, Utrecht, the Netherlands\\ \email{{v.steindani,x.lu,i.m.beerepoot}@uu.nl}}

\maketitle              
\begin{abstract}
In this short paper, we explore the enrichment of event logs with data from wearable devices. We discuss three approaches: (1) treating wearable data as event attributes, linking them directly to individual events, (2) treating wearable data as case attributes, using aggregated day-level scores, and (3) introducing new events derived from wearable data, such as sleep episodes or physical activities. To illustrate these approaches, we use real-world data from one person, matching health data from a smartwatch with events extracted from a digital calendar application. Finally, we discuss the technical and conceptual challenges involved in integrating wearable data into process mining for personal productivity and well-being.

\keywords{Event log enrichment, Wearable devices, Well-being}
\end{abstract}
%
%
%
\section{Introduction} 
Imagine a knowledge worker whose calendar records a day of back-to-back virtual meetings, a focused work session, and a late-afternoon presentation. While the schedule captures the sequence of activities, it misses the lived experience: the rising stress before the presentation, the prolonged inactivity of sitting through multiple calls, or the restorative effect of a short walk during lunch. This information is crucial for understanding how work practices affect personal well-being and the other way around.  

Process mining traditionally relies on event logs generated by information systems, with a primary focus on efficiency and compliance~\cite{Aalst2016}. However, this approach often overlooks the human side of processes. A growing interest in human-centric process mining seeks to address this gap by incorporating data on individual well-being, behavior, and context~\cite{mannhardt2018privacy,braakman2024mining}. Recent advances in wearable technology offer new possibilities for this endeavor. Smartwatches, fitness trackers, and sleep monitors provide continuous streams of physiological data, such as heart rate variability (HRV), physical activity, and sleep quality~\cite{Seneviratne2017, Brzychczy2025}. When combined with organizational data such as calendar logs, these signals enable richer analyses of how daily activities interact with health and performance. This integration opens novel lines of inquiry, such as:
\begin{itemize}
    \item How do different meetings correlate with indicators of stress or recovery?
    \item To what extent does physical activity coincide with productive workdays?
    \item How does sleep quality influence the execution of daily work processes?
    \item How do meeting-heavy days affect subsequent sleep and recovery patterns?
    \item At what times of day do demanding tasks align with physiological readiness?
\end{itemize}

In this paper, we propose and illustrate three complementary strategies for integrating wearable data into personal process mining. By connecting scheduled activities with physiological signals, our approach empowers knowledge workers to better understand how they work and establish healthier routines.

\section{Approach and Illustration}
Our approach builds on the integration of event data with physiological and behavioral data from wearable devices and distinguishes three complementary strategies for integrating wearable data into process mining. In our illustration, calendar events are extracted from Microsoft Outlook, while metrics such as HRV, sleep episodes and physical activities are obtained from an XML export of an Apple Watch. Our study participant shared over 8 months worth of data, from 13 November 2024 to 26 July 2025. 
Next, we describe the different strategies for enriching organizational event data with wearable data, illustrating them with real-world~data. To do so, we processed and analyzed the data in a Jupyter Notebook\footnote{Notebook can be found here: \url{https://github.com/ibeerepoot/wearable_pm}} and used Fluxicon's Disco\footnote{\url{https://fluxicon.com/disco/}} to analyze the event log, before discussing the results with the participant.  

\mypar{Treating Wearable Data as Event Attributes}
Firstly, measurements from wearables can be matched directly to the time window of events. To demonstrate this strategy, we aligned HRV measurements (commonly used for measuring stress~\cite{kim2018stress}) with calendar events extracted from Microsoft Outlook. The Apple Watch measures HRV throughout the day and night, using the standard deviation of NN (SDNN) metric. We parsed the SDNN measurements from Apple Health's XML exports and the Outlook events from the CSV file, filtering out all-day events. Next, we matched each Outlook event with (the median of) HRV measurements recorded during its time window. Of the 452 calendar events, 314 were matched with one or more HRV measurements. 

\begin{figure}[!t]
    \centering
    \begin{minipage}{0.48\linewidth}
        \centering
        \includegraphics[width=\linewidth]{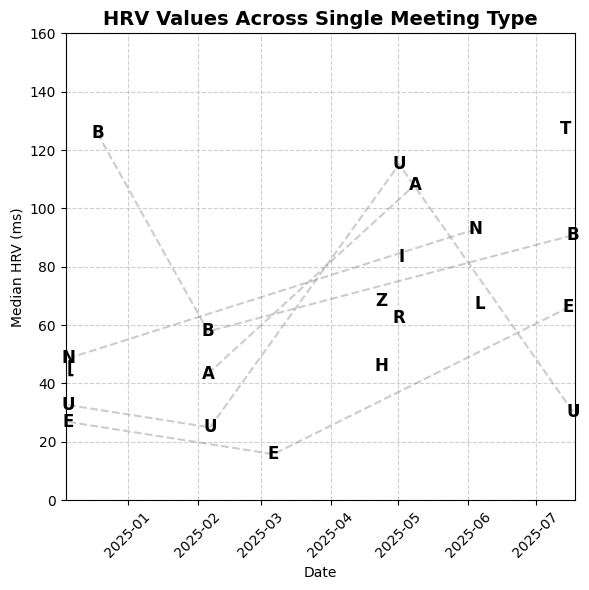}
        \caption{Depiction of HRV values across internal project meetings (marked with project leader's initial).}
        \label{fig:ipos}
    \end{minipage}\hfill
    \begin{minipage}{0.48\linewidth}
        \centering
        \includegraphics[width=\linewidth]{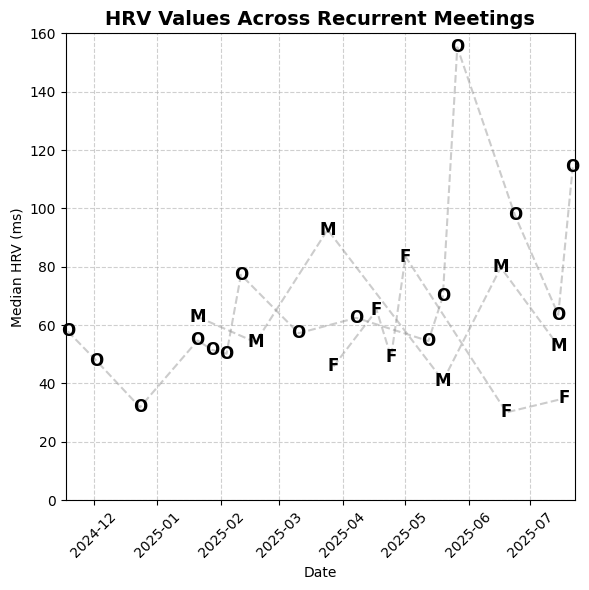}
        \caption{Depiction of HRV values across three types of recurrent meetings (marked with first letter of meeting).}
        \label{fig:recurrent}
    \end{minipage}
\end{figure}

In Figure \ref{fig:ipos}, we illustrate one type of meeting that the participant holds frequently, namely internal project meetings with project leaders (each labeled by their initial). This view allows the participant to analyze whether there are meetings with project leaders that are more or less stressful, and spot exceptionally stressful ones. Generally, internal project meetings with colleagues U and E show low (negative) HRV values, although one meeting with colleague U in May 2025 was paired with an exceptionally high HRV value (115ms). These internal project meetings, however, are held at different times of day, on different days of the week, which could influence HRV values and make comparison difficult. 
In Figure \ref{fig:recurrent}, we take a different perspective and illustrate three recurrent meetings that take place regularly, at the same time and on the same day. While the participant identified meeting M as the more stressful and meetings O and F as more lightweight meetings, this is not clearly reflected in the data. 

\mypar{Treating Wearable Data as Case Attributes}
Wearable devices typically provide health data aggregated at the day level, e.g., a daily sleep score. Device-specific aggregate scores include Garmin's body battery, Oura's readiness, and Fitbit's stress score. While these aggregate scores cannot be directly linked to individual events of a workday, they can be stored as case attributes for analyzing event sequence impacts. To demonstrate this, we extracted daily heart rate and sleep data from the study participant's Apple Health export. Resting heart rate is available at the day level, while sleep data is stored as episodes of different sleep stages and time awake, which can be aggregated in post-processing. 
For the purpose of this illustration, we stored resting heart rate, total sleep time, total time awake, and total time in deep sleep. We included the sleep episodes at the end of a workday to analyze how workday events relate to subsequent sleep quality, though reverse analysis is also possible. 

\begin{figure}[!t]
    \centering
    \includegraphics[width=0.95\linewidth]{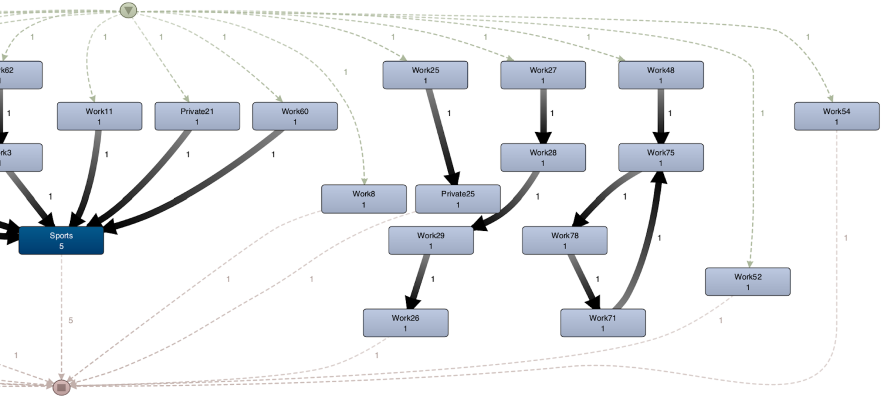}
    \caption{Section of the discovered process model for workdays with a good night's sleep.}
    \label{fig:disco}
\end{figure}

Figure~\ref{fig:disco} shows a cohort of the discovered process model for workdays that were followed by a good night's sleep, specifically, 8 hours or more of total sleep time and less than 1 hour of awake time (Outlook event names were renamed to preserve privacy). Generally, these days seem to be characterized by the inclusion of sports, a dinner (Private25), a work lunch meeting (Work26, Work71, Work54), or social drinks at work (Work52). 

\begin{figure}[h!]
    \centering
    \begin{minipage}{0.63\linewidth}
        \centering
        \includegraphics[width=\linewidth]{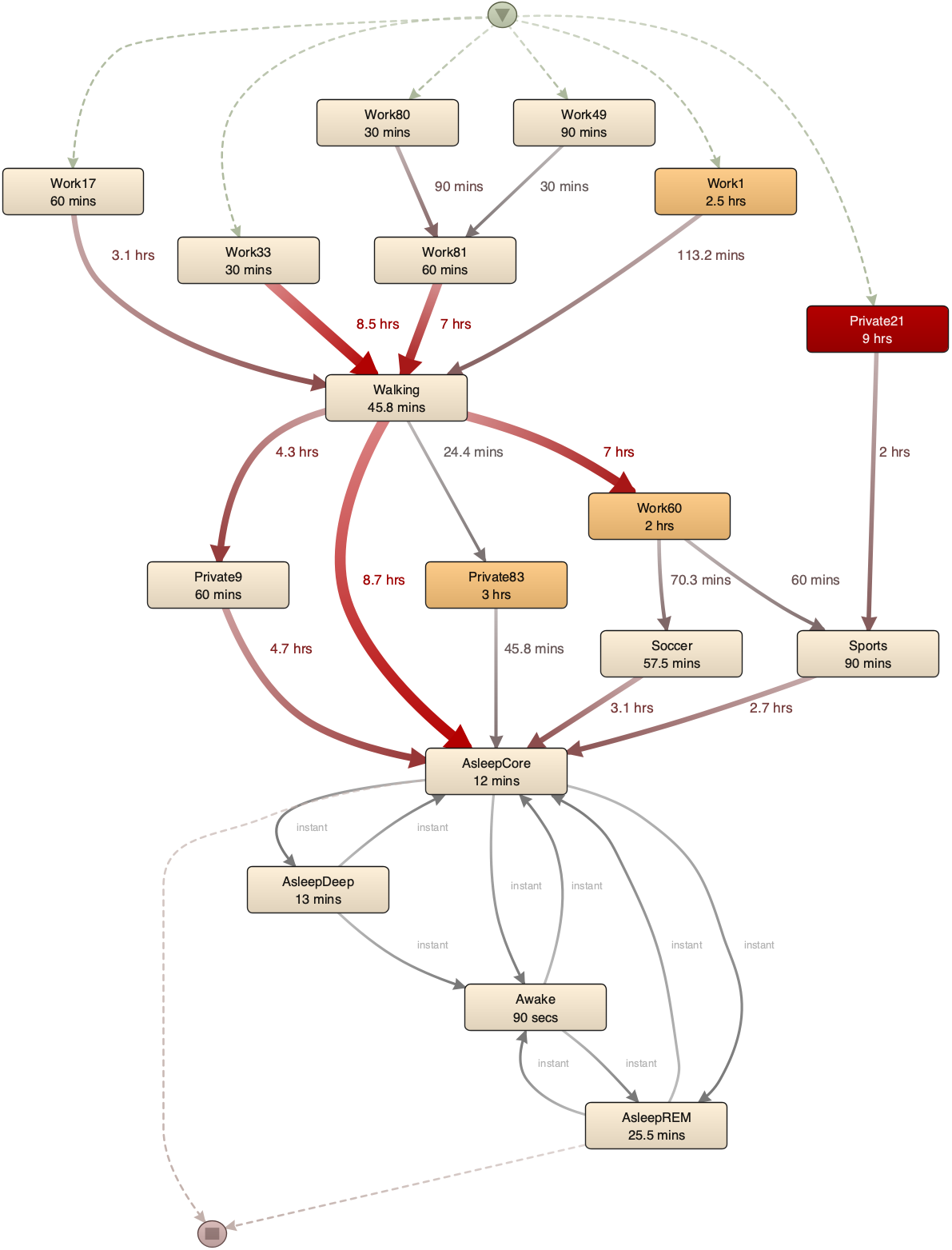}
        \caption{Median duration projected onto discovered process model for workdays that include walking.}
        \label{fig:walk_and_work}
    \end{minipage}\hfill
    \begin{minipage}{0.3\linewidth}
        \centering
        \includegraphics[width=\linewidth]{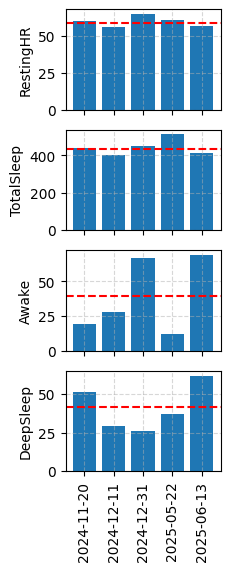}
        \caption{Case attribute values for cases shown in Fig. \ref{fig:walk_and_work} (dashed red line shows average values across all workdays).}
        \label{fig:walking_days}
    \end{minipage}
\end{figure}

\mypar{Introducing New Events Derived from Wearable Data}
Beyond enriching existing events and cases, wearable data enables the introduction of entirely new event types into an event log. In our study, we incorporated the participant's workouts and sleep episodes as additional events. Figure~\ref{fig:walk_and_work} illustrates the discovered process model for workdays during which the Apple Watch detected walking activity, while Figure~\ref{fig:walking_days} presents the corresponding case attribute values. These walking days appear to be associated with average levels of resting heart rate (RestingHR) and total sleep duration (TotalSleep). Notably, three of these days also show minimal time spent awake (Awake), although two outliers diverge from this pattern: one coinciding with New Year’s Eve and the other following an engaging movie that may have prolonged time awake. 

\section{Conclusion and Challenges}
While enriching event logs with wearable data holds promise for personal and human-centric process mining, it also presents several key challenges:

\mypar{1. Selecting appropriate indicators of well-being} 
There is ongoing debate about how best to measure constructs like stress and recovery. Heart rate variability (HRV) is commonly used as a proxy for physiological stress, but it has known limitations and may not reliably capture mental or emotional strain in all contexts. While more direct or clinically validated measures exist, they are often intrusive or impractical for continuous, real-world monitoring.
    
\mypar{2. Disentangling cause, effect, and confounding factors} 
Even if we assume a reliable well-being indicator, interpreting it in context remains difficult. For example, elevated stress during a meeting may reflect the meeting itself, but it could just as well be influenced by poor sleep the night before, a skipped workout, or unrelated personal circumstances. Enriched logs risk oversimplifying complex human states unless multi-factor interactions are carefully considered.

\mypar{3. Ensuring ethical use and data protection}
Wearable-derived data can reveal sensitive health-related information, making it critical to enforce safeguards such as data minimization and personal ownership when integrating physiological and behavioral data. We suggest limiting such analyses to personal usage, where a worker can choose to share aggregated findings should they deem this valuable. 

\bibliographystyle{splncs}
\bibliography{ref}

\end{document}